**Functional neuroimaging of psychedelic experience: An overview of psychological and neural effects and their relevance to research on creativity, daydreaming, and dreaming**

Kieran C. R. Fox[a], Manesh Girn[a], Cameron C. Parro[a], and Kalina Christoff[a,b]

[a] Department of Psychology, University of British Columbia, 2136 West Mall, Vancouver, B.C., V6T 1Z4 Canada

[b] Brain Research Centre, University of British Columbia, 2211 Wesbrook Mall, Vancouver, B.C., V6T 2B5 Canada

*Corresponding author:* Fox, K.C.R. (kfox@psych.ubc.ca)

To appear in: *The Cambridge Handbook of the Neuroscience of Creativity*. (R.E. Jung & O. Vartanian, Editors). Cambridge University Press.

**Introduction: A very brief history of psychedelic substances and science**

Humans have employed an incredible variety of plant-derived substances over the millennia in order to alter consciousness and perception (Schultes, Hofmann, & Rätsch, 2001). Among the innumerable narcotics, analgesics, 'ordeal' drugs, and other psychoactive substances discovered and used in ritualistic contexts by cultures around the world, one class in particular stands out not only for its radical psychological effects, but also for the highly charged political and legal atmosphere that has surrounded it since its widespread adoption about 50 years ago: so-called psychedelic substances.

In the 1950's and 1960's, psychedelic substances that had been used for thousands of years in indigenous communities, particularly in Mesoamerica and the Amazon basin, were rediscovered by adventurous foreigners and introduced into Western societies that had been largely ignorant of the use of such substances for centuries, if not millennia (McKenna, Towers, & Abbott, 1984; Schultes, 1957; Wasson, 1958; Wasson, Hofmann, Ruck, & Smith, 2008). These substances soon spilled beyond the boundaries of traditional use (Tupper, 2008) and were rapidly adopted by the countercultural movement of the time (J. Stevens, 1987). Around the same time as this Western adoption of natural psychedelic substances, synthetic and semi-synthetic psychedelics were being discovered or re-discovered in research laboratories, most notably lysergic acid diethylamide (LSD) in 1943 (Hofmann, 1980) and ketamine in 1962 (C. L. Stevens, 1966).

Interest in these substances was not limited to the counterculture movement: scientists, psychiatrists, and others saw potential for psychological research (Lilly, 1972), enhancement of creative thinking (Harman, McKim, Mogar, Fadiman, & Stolaroff, 1966) and artistic appreciation (Huxley, 1954), and clinical value in psychotherapy (Pahnke, Kurland,

Unger, Savage, & Grof, 1970) and the treatment of addiction (Krebs & Johansen, 2012). Military and espionage applications were also envisioned, leading to a now well-documented series of non-consensual (and illegal) misadventures in psychedelic 'research' by the Central Intelligence Agency in the United States (Lee & Shlain, 1992) and MI6 in the United Kingdom (Streatfeild, 2008).

Essentially all such substances were criminalized on an international scale in the wake of the United Nations' Single Convention on Narcotic Drugs (1961) and Convention on Psychotropic Substances (1971) (McAllister, 2000). With criminalization and its concomitant social stigma, research on psychedelics in humans effectively ceased for several decades. Given that many promising avenues of research were revealed by the early investigations of the 1950's and 1960's, this hiatus very likely hindered progress in the fields of neuroscience, psychology, and psychiatry (Nutt, 2014; Nutt, King, & Nichols, 2013).

Even as psychedelic research stalled, the neuroscientific tools available for understanding human brain function were advancing by leaps and bounds. Rapid technological developments led to remarkable improvements in the ability to image the metabolism and functioning of the human brain non-invasively and at high spatial and temporal resolutions, in particular with positron emission tomography (PET) and functional magnetic resonance imaging (fMRI) (Huettel, Song, & McCarthy, 2004; Raichle, 2009; Savoy, 2001). These functional neuroimaging modalities were rapidly applied in the investigation of the neural bases of many uniquely (or at least largely) human cognitive capacities that were difficult or impossible to study in animal models, including language processing (McCarthy, Blamire, Rothman, Gruetter, & Shulman, 1993; Petersen, Fox,

Posner, Mintun, & Raichle, 1988), visual imagery (Kosslyn, Thompson, & Alpert, 1997), creativity (Bekhtereva et al., 2000), dreaming (Maquet et al., 1996), meditation (Lazar et al., 2000; Lou et al., 1999), empathy (Farrow et al., 2001), and mentalizing (Fletcher et al., 1995) – all of which have since developed into burgeoning sub-fields with dedicated journals, conferences, and research groups, and hundreds, if not thousands, of investigations of their neural correlates. The first PET study of psychedelic experience was similarly reported early on (Vollenweider, Leenders, Scharfetter, Antonini, et al., 1997), but in contrast, relatively little functional neuroimaging research has followed over the subsequent two decades (see Table 1 for a list of functional neuroimaging studies of psychedelic experience).

Despite various practical, political, and legal hurdles, however, recent years have seen an acceleration of the timid renaissance in human psychedelic research (Kupferschmidt, 2014; Langlitz, 2007; Sessa, 2012b), paralleled by growing evidence that psychedelics are among the least-harmful and least-addictive pharmacological substances regularly used by humans (Nichols, 2004; Nutt, King, & Phillips, 2010; Nutt, King, Saulsbury, & Blakemore, 2007). The last few years have witnessed an increasing number of functional neuroimaging investigations of 'classic' psychedelic substances such as psilocybin (Carhart-Harris, Erritzoe, et al., 2012; Carhart-Harris et al., 2013; Carhart-Harris, Leech, et al., 2012; Lebedev et al., 2015; Roseman, Leech, Feilding, Nutt, & Carhart-Harris, 2014; Vollenweider, Leenders, Scharfetter, Maguire, et al., 1997) and ayahuasca (Bouso et al., 2015; de Araujo et al., 2012; Palhano-Fontes et al., 2015; Riba et al., 2006), as well as substances with related effects, such as ketamine (Grimm et al., 2015; Pollak et al., 2015; Vollenweider, Leenders, Scharfetter, Antonini, et al., 1997). Additionally, a recent study has finally offered the first investigation of the neural correlates of the LSD state (Carhart-Harris et al., 2016). We summarize this research in Table 1, including the general stage(s) of the psychedelic experience each study investigated (note that we primarily include studies that investigated neural correlates of the psychedelic experience *per se*, but omit studies that investigated tasks performed under the influence of psychedelic substances).

This burgeoning functional neuroimaging research is beginning to provide long-sought answers to questions about the neural correlates of psychedelic experiences. The purpose of this chapter is to review this small but growing body of functional neuroimaging research, and address two key questions. The first is whether differential neural correlates accompany the various stages of psychedelic experience: for instance, are there specific neural correlates accompanying the initial transition from baseline; the core psychedelic experience; and so-called 'peak' experiences? Second, although the psychedelic state has most often been compared with the psychological extremes of psychosis (Gonzalez-Maeso & Sealfon, 2009) or profound religious experience (Pahnke, 1969), we believe that much can be gained from a comparison with kindred but naturally-occurring altered states of consciousness, specifically dreaming, daydreaming, and creative thinking – all of which bear similarities to the psychedelic experience in terms of their general tendency toward internally-focused attention, their potential for insightful or novel patterns of thinking, and the greater or lesser prevalence of immersive, imagined sensory experience (Dittrich, 1998).

*Table 1.*  Neuroimaging investigations of psychedelic experience.

| Study | Modality | Substance | Stage(s) Investigated[a] |
|---|---|---|---|
| Vollenweider et al. (1997) | PET | Psilocybin | Peak |
| Holcomb et al. (2001) | PET | Ketamine | Onset, Core, Peak, Resolution |
| Långsjö et al. (2003) | PET | Ketamine | Core |
| Aalto et al. (2005) | PET | Ketamine | Core |
| Deakin et al. (2008) | fMRI | Ketamine | Onset |
| Carhart-Harris et al. (2012a) | fMRI | Psilocybin | Onset, Core |
| Carhart-Harris et al. (2012b) | fMRI | Psilocybin | Onset, Core |
| De Araujo et al. (2012) | fMRI | Ayahuasca | Core, Peak |
| Scheidegger et al. (2012) | fMRI | Ketamine | Core, Assimilation |
| Carhart-Harris et al. (2013) | fMRI | Psilocybin | Onset, Core |
| De Simoni et al. (2013) | fMRI | Ketamine | Onset |
| Driesen et al. (2013) | fMRI | Ketamine | Onset, Core |
| Roseman et al. (2014) | fMRI | Psilocybin | Onset, Core |
| Bouso et al. (2015) | Morphometric MRI | Ayahuasca | Sequelae |
| Grimm et al. (2015) | fMRI | Ketamine | Core |
| Joules et al. (2015) | fMRI | Ketamine | Onset, Core |
| Khalili-Mahani et al. (2015) | fMRI | Ketamine | Core, Peak, Resolution |
| Lebedev et al. (2015) | fMRI | Psilocybin | Onset, Core |
| Palhano-Fontes et al. (2015) | fMRI | Ayahuasca | Core |
| Pollak et al. (2015) | ASL (MRI) | Ketamine | Onset |
| Carhart-Harris et al. (2016) | ASL, fMRI, MEG | LSD | Onset, Core, Peak |

*Notes.* [a] Refer to our classification scheme in Table 3.
ASL: arterial spin labeling; fMRI: functional magnetic resonance imaging; MEG: magnetoencephalography; MRI: magnetic resonance imaging; PET: positron emission tomography.

**Overview of the major psychedelic substances**

There are about a half-dozen major psychedelic substances in widespread use (Table 2), but dozens, if not hundreds, of others have been identified in nature (Schultes et al., 2001) or synthesized in laboratories (Shulgin & Shulgin, 1997). The neurochemical mechanisms at the level of the synapse (i.e., pharmacokinetics and pharmacodynamics) are understood in broad outline for the widely-used substances (Table 2), but the precise details are highly complex, poorly understood, and based almost exclusively on animal models (Nichols, 2004). The most salient finding to date is that many 'classic' psychedelics appear to act primarily via the serotonin system, as agonists of the 5-HT$_{2A}$ receptor in particular (Glennon, Titeler, & McKenney, 1984; Nichols, 2004). The structural simplicity of these substances at the molecular level (small in size and largely planar; Table 2) belies the profound influence they have on perception, memory, emotion, and sense of self, even at extremely low doses (in the microgram range in some cases).

Ketamine is not generally considered a classic psychedelic substance, but is instead typically classified as a 'dissociative anesthetic,' and has a distinct molecular mechanism of action (Table 2). Nonetheless, it can induce many experiences similar to those associated with 'classic' psychedelics (Studerus, Gamma, & Vollenweider, 2010; Vollenweider & Kometer, 2010), and moreover (to anticipate our results somewhat) the large-scale neural activity engendered by ketamine appears to be appreciably similar to that elicited by serotonin agonists such as psilocybin. Ketamine is also the best-studied psychedelic substance in terms of the number and variety of functional neuroimaging studies conducted to date. We therefore include ketamine in our discussions here.

*Table 2.*  Overview of the major psychedelic substances.

| Substance (common name) | Main active constituent | Neurochemical mechanisms of action (main receptor affinity) | Chemical structure | Key references |
|---|---|---|---|---|
| LSD / Acid | Lysergic acid diethylamide | Serotonin agonist (esp. 5-HT$_{2A}$) Dopamine (all subtypes) Adrenoceptors (all subtypes) | 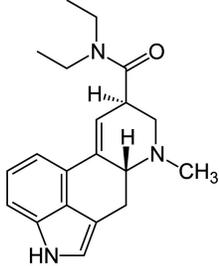 | (Marona-Lewicka, Thisted, & Nichols, 2005; Nichols, 2004) |
| Magic mushrooms | Psilocybin | Serotonin (esp. 5-HT$_{2A}$; also 5-HT$_{1A}$, 5-HT$_{1D}$, 5-HT$_{2C}$) | 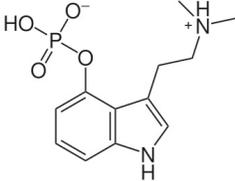 | (Passie, Seifert, Schneider, & Emrich, 2002) |
| Ayahuasca / DMT | *N,N*-Dimethyltryptamine | Serotonin agonist (esp. 5-HT$_{2A}$; also 5-HT$_{1A}$, 5-HT$_{1B}$, 5-HT$_{1D}$, 5-HT$_{2B}$, 5-HT$_{2C}$) | 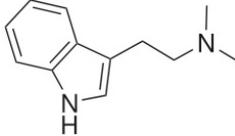 | (Deliganis, Pierce, & Peroutka, 1991; Keiser et al., 2009; Ray, 2010) |
| Peyote / Mescaline | 3,4,5-Trimethoxyphenethylamine | Serotonin agonist (esp. 5-HT$_{2A}$ and 5-HT$_{2C}$) | 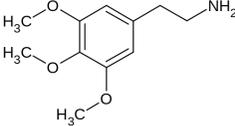 | (Monte et al., 1997; Nichols, 2004) |
| Special K | Ketamine | NMDA antagonist Opioid agonist | 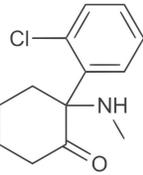 | (Brockmeyer & Kendig, 1995; Editorial, 1996; Jansen & Sferios, 2001; Kohrs & Durieux, 1998; Salt, Wilson, & Prasad, 1988) |
| Salvia divinorum | Salvinorin A | κ-Opioid agonist | 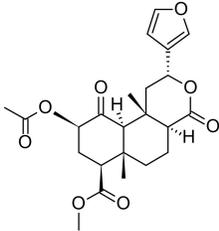 | (Roth et al., 2002) |

**The psychedelic experience and its relationship to creativity, daydreaming, and dreaming**

The psychedelic experience is by its very nature highly creative, often involving generation of a high volume of novel ideas and insights; profuse visual, auditory, and somaesthetic hallucinations; and intense, widely-valenced emotional experiences (Dittrich, 1998). These profound alterations in consciousness have most often been compared with either psychosis on the negative end of the spectrum (Vollenweider & Kometer, 2010) or transcendent religious and mystical experiences on the positive end of the spectrum (Pahnke, 1967, 1969). Less sensational comparisons, however, can be made with the psychological experiences and neural correlates of more mundane 'altered' states, including daydreaming, nighttime dreaming, and creative thinking. The main features of overlap are immersive visionary/hallucinated experiences (as in dreaming and, to a lesser extent, daydreaming); the loss of a sense of self (in peak experiences of creative thinking and inspiration, and to a lesser extent in dreaming as well); and in the need (in conscientious users, at least) to assess and evaluate the altered state experience in later periods of assimilation (as is common in creative idea generation, and in dreaming).

*Relationship to artistic, scientific, and philosophical creation*

Artists and other creative individuals have often reported using psychedelic substances in an effort to enhance creative output or novelty (Sessa, 2008), and some early experimental work suggested positive effects along these lines (Harman et al., 1966; Krippner, 1972). There are even reports (albeit anecdotal) of psychedelic experiences being centrally involved in major scientific breakthroughs, such as Kary Mullis's Nobel prize-winning discovery of the polymerase chain reaction (Mullis, 2010).

Although descriptions of the creative process tend to be more muted today, historically, many artists and poets have described entering an altered state of consciousness reminiscent of psychedelic experience (Kivy, 2001; McMahon, 2013; Murray, 1989). Indeed, Plato's theory of creative inspiration directly credited the possession of the artist by the Muse or other gods; the artist lost their own mind or self and entered a poetic 'fury' or 'frenzy' during which a higher force created *through* them (McMahon, 2013; Murray, 1996; Pieper, 1964). Plato even went so far as to directly compare the frenzy of artistic creation with the cultic rites of Dionysus, where participants were similarly filled with (probably substance-induced) ecstasy and enthusiasm (Murray, 1996). William Blake provides an excellent (if more modern) example: Blake's poetry and painting are all but unimaginable absent the visionary altered states that inspired them (Bloom, 1963). The loss of the sense of self in visionary states is in many ways reminiscent of core and peak psychedelic experiences, discussed in more detail below.

Similarly, although philosophy today brings to mind dry academic discussions and impenetrable texts, the search for philosophical truth in Ancient times strongly endorsed altered states of consciousness as a valid source of knowledge and understanding. The timeless mystical vision of Plato's Allegory of the Cave (Plato, Grube, & Plochmann, 1974) is probably the most well-known example, but many other Classical philosophers made visionary, mystical experience the foundation upon which they later built elaborate philosophical systems of their own creation, including Plotinus (Plotinus & Katz, 1950; Plotinus & MacKenna, 1969) and likely also Pythagoras (Burkert, 1972). Mystical experience has similarly played a fundamental role in the rich history of Eastern philosophy and spirituality, as especially apparent in Buddhist, Hindu, and Jain traditions (Sarma, 2011). In the fine arts, the sciences, and even philosophy, then, the notion that visionary and ego-dissolving altered states are a legitimate (perhaps even necessary) ground upon which to base creative output has a long history.

Another important parallel is that many theories of creativity propose a 'two-stage' model, whereby initially generated ideas must then be subsequently evaluated for novelty and utility (Basadur, Graen, & Green, 1982; Beaty, Benedek, Silvia, & Schacter, 2015; Ellamil, Dobson, Beeman, & Christoff, 2012). We suggest that, similarly, an 'ideal' psychedelic experience is not simply accepted uncritically, no matter how profound its insights might be, but is later evaluated and assimilated into prior knowledge structures, even as it may expand the boundaries or alter the form of these very structures (we call this the *assimilation* stage; see Fig. 1 and Table 3). Another important similarity, then, is the need to subsequently evaluate the highly novel and creative experiences generated in the psychedelic state. The same principle is applied by those who pay close attention to dream experience: dreams are typically analyzed, interpreted, and assessed for significance and insights after the fact.

*Relationship to daydreaming and dreaming.*

Although the vividness and content of daydreaming are not widely appreciated, detailed experience sampling has revealed that mild-to-moderate visual imagery and fantasy are ubiquitous throughout our everyday lives (Fox, Nijeboer, Solomonova, Domhoff, & Christoff, 2013; Klinger, 2008; Klinger & Cox, 1987; Stawarczyk, Majerus, Maj,

Van der Linden, & D'Argembeau, 2011). Dreaming is a more familiar example of a bizarre and visionary experience that each of us participates in every night; there are many parallels with intense psychedelic experience, including immersive visual imagery, highly novel and unusual patterns of thinking, intensive emotional experience, and a disrupted sense of self (Fox et al., 2013; Windt, 2010, 2015). Specific examples of parallel brain recruitment and subjective experience will be discussed as they arise throughout the remainder of this chapter.

**Stages of psychedelic experience: Phenomenology and neural correlates**

Users of psychedelic substances have long reported that a typical psychedelic experience is by no means uniform, but rather follows a general trajectory (Fig. 1). It is tempting to simply map this experiential trajectory onto the putative concentrations of active substance in the blood and brain, but such a view fails on several counts. Research *does* suggest that various dimensions of psychedelic experiential intensity indeed increase in a dose-dependent fashion (Bowdle et al., 1998; Studerus, Kometer, Hasler, & Vollenweider, 2011; Vollenweider & Kometer, 2010), and so substance concentration may well follow a broadly similar trajectory in many cases. Nonetheless, a monotonic relationship between substance concentration and experiential intensity cannot explain the enormous influence of set (i.e., psychological expectations and cognitive-affective state of the individual) and setting (i.e., factors including physical environment, social milieu, and so on) (Nichols, 2004; Nour & Krzanowski, 2015; Sessa, 2012a). Moreover, it is increasingly well-documented that only some features of experiential intensity parallel substance concentration (Studerus et al., 2011): psychological and neural changes persist long after the acute phase of the psychedelic experience (Bouso et al., 2015; Griffiths et al., 2011; Griffiths, Richards, McCann, & Jesse, 2006; Studerus et al., 2011), and long after substance concentrations in the blood and brain are negligible or zero. On the psychological side, subsequent reflection upon, and evaluation of, the acute psychedelic experience, as well as alterations in mood and personality, may persist indefinitely, perhaps becoming less marked over time, but never diminishing entirely – i.e., diminishing asymptotically (Fig. 1).

*Table 3.*  Phases and stages of psychedelic experience.

| Phase | Stage | Overview |
|---|---|---|
| Acute (0.5 hrs – 12 hrs) | Onset | Initial transition into the psychedelic state following ingestion of a substance |
| | Core | Sustained effects of the core psychedelic experience, often strongly visual in nature |
| | Peak | 'Peak' experiences of ego-dissolution, mystical union |
| | Resolution | Denouement and 'come-down'; the cessation of acute effects |
| Sub-acute (hours – days) | Assimilation | Lingering effects and insights immediately following the experience; evaluation and contemplation of acute experience |
| Long-term (days – years) | Sequelae | Long-term effects on brain structure and function, personality, worldview, and other cognitive-affective variables |

*Notes.* Various putative phases and stages of a typical psychedelic experience. The grey shading of the 'peak' stage indicates that this stage may be much more rare and vary enormously in intensity as compared to the other stages.

*Figure 1.* An idealized representation of the phases and stages of psychedelic experience.

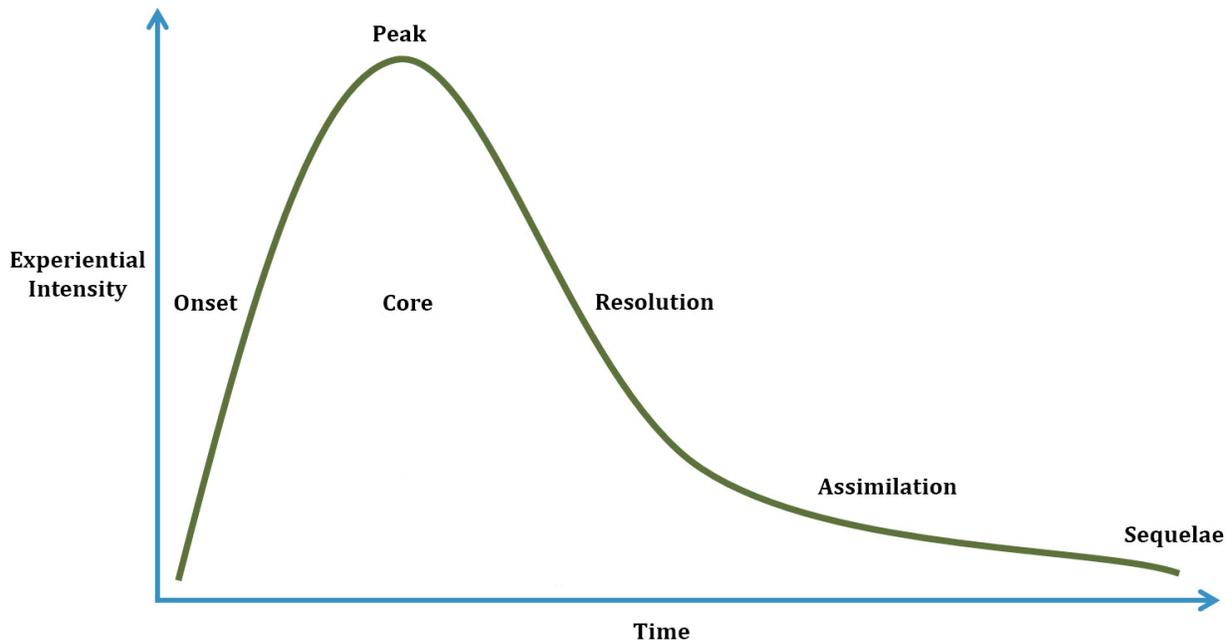

Our model of psychedelic phases and stages (Fig. 1; Table 3) is meant merely as a crude preliminary scheme for beginning to parse and understand the growing body of functional neuroimaging evidence. Aside from the huge variability to be expected at the individual level, we hasten to underline the fact that our smooth curve of experiential intensity is merely an abstraction or idealization that does not take into account the complexity of a typical psychedelic experience, which can unfold unpredictably and with varying intensity. We therefore intend the model not as a hypothesis to be tested, but merely as a helpful classification scheme.

*Onset*

The *onset* stage corresponds to the initial transitional period following drug intake, during which subjective effects first start to take hold. Depending on a variety of factors, including the particular substance ingested and route of administration used, this stage can vary anywhere from as little as 30 seconds to over 60 minutes. Phenomenologically, this stage involves the initial changes in perception, cognition, internal thought, and sense of self. General physiological arousal often rises, resulting in some level of anxiety, increased heart-rate, and sometimes visual hallucinatory effects, in the case of classic psychedelics such as psilocybin and LSD (Hollister, 1984). With ketamine, on the other hand, the initial transition stage tends to be accompanied by impaired thought and a sense of dissociation from the body (Jansen & Sferios, 2001).

So far, only work with ketamine has directly investigated the neural correlates of this early stage of the psychedelic experience. In one study, Deakin and colleagues (2008) intravenously administered ketamine to 12 participants and observed minute-to-minute brain activity changes over a time-course from 8 minutes pre- to 8 minutes post-infusion using fMRI. The authors observed rapid onset, with peak subjective effects at 4-5 minutes following infusion (Deakin et al., 2008). The first significant neural effects observed were deactivations in medial orbitofrontal cortex and the temporal pole two minutes post-infusion (T2). Deactivation then spread to the subgenual anterior cingulate and frontopolar cortex at three-minutes post-infusion (T3).

The T2 deactivations in the medial prefrontal cortex and the temporal pole were highly correlated with self-reported scores on 'derealization' and 'depersonalization' (Deakin et al., 2008). Additionally, medial orbitofrontal cortex deactivation was uniquely correlated with self-reported 'thought disorder' and 'hallucinations' (Deakin et al., 2008). These results are intriguing, because medial prefrontal cortex and temporopolar cortex are both key regions supporting a sense of self and internal streams of thought, such as mind-wandering, daydreaming, and other kinds of self-referential thinking (Andrews-Hanna,

Smallwood, & Spreng, 2014; Ellamil et al., 2016; Fox, Spreng, Ellamil, Andrews-Hanna, & Christoff, 2015).

Overall, in line with the subjectively reported dissociative effects of ketamine, fMRI investigations suggest a deactivation of key midline brain areas involved in sustaining a coherent stream of thought and sense of self during the onset of ketamine-induced experiences. Intriguingly, these results parallel findings of more serious disruptions of the default network being associated with full-blown ego-dissolution in peak experiences (see below). The interested reader should consult another study investigating ketamine onset by Driesen and colleagues (2013).

*Core experience*

The *core* stage of our model pertains to the bulk of a psychedelic experience, and can last anywhere from 10 minutes to upwards of 8 hours. The core stage has been explored in considerably more research, including independent work on psilocybin (Carhart-Harris, Erritzoe, et al., 2012), ayahuasca/DMT (Palhano-Fontes et al., 2015), ketamine (Långsjö et al., 2003), and most recently, LSD (Carhart-Harris et al., 2016) (refer to Table 1 for a full list of studies investigating the core stage). Although the core psychedelic experience can, of course, involve a nearly infinite range of effects, one of the most prevalent and predictable experiences, especially at high doses, is prominent visual imagery, illusions, and hallucinations. We therefore focus our discussion of the core experience on this ubiquitous feature of the psychedelic state.

One recent study investigated core experiences following intravenous injection of LSD using numerous noninvasive neuroimaging modalities (Carhart-Harris et al., 2016). A central result was increased cerebral blood flow and resting-state functional connectivity in visual cortex areas during the core LSD experience, and the finding that self-reported visual illusions and hallucinations correlated significantly with this altered activation in visual areas (Carhart-Harris et al., 2016). These results are intriguing because medial visual cortex areas are consistently activated during waking mind-wandering (Fox et al., 2015), which involves ubiquitous visual imagery (Fox et al., 2013; Klinger & Cox, 1987; Stawarczyk et al., 2011), as well as during REM sleep and dreaming – likewise characterized by vivid visual imagery and hallucinations (Domhoff & Fox, 2015; Fox et al., 2013; Schredl, 2010). Indeed, the cluster of increased visual cortex activation observed during the LSD experience overlaps strikingly with the clusters found in meta-analyses we conducted of the neural correlates of waking daydreaming (Fox et al., 2015) and nighttime dreaming (Fox et al., 2013) – presumably because this area is subserving the same function in each case (compare Fig. 1 in Carhart-Harris et al., 2016, with Fig. 2c in Fox et al., 2015 and Fig. 1i in Domhoff & Fox, 2015). Note that with LSD this visual cortex activation was observed compared to a resting placebo-controlled baseline, during which (presumably) there was a large degree of spontaneously generated thought and imagery. These results suggest that the same visual area apparently involved in normal waking visual imagery is more highly activated when visual illusions and hallucinations become more vivid and frequent, as during a typical psychedelic experience. The studies reporting this medial visual activation in REM sleep (Fox et al., 2013) likewise compared brain activity to a resting, waking baseline, leading us to a similar interpretation: the increased vividness and hallucinatory quality of dreams appears to more strongly recruit the same visual areas active in waking rest – just as does visionary psychedelic experience.

Another recent study explored the visionary effects of the brew ayahuasca (de Araujo et al., 2012), whose central active ingredient is dimethyltryptamine, another serotonin receptor agonist (see Table 2). Similar to the LSD results above, the authors observed widespread activation of visual cortex areas during an eyes-closed imagery task after ayahuasca consumption (de Araujo et al., 2012), including regions such as the lingual gyrus (BA 17) that are activated during daydreaming (Fox et al., 2015) and nighttime dreaming (Domhoff & Fox, 2015; Fox et al., 2013).

Of course many other subjective experiences and neurophysiological findings have been reported during core experiences for various substances and under various conditions. We limit our discussion here for the sake of brevity, but refer the reader to Table 1 for further studies of core psychedelic experience.

*Peak experience: Ego-dissolution and mystical unity*

The *peak* stage of a psychedelic experience is the period in which subjective effects reach their maximum intensity (Fig. 1). Although some peak of intensity is presupposed by the rise and fall of effects over the course of the psychedelic experience (Fig. 1), peak experiences can vary enormously in intensity and duration, from a mere amplification of core effects to qualitatively distinct 'dissolution' and 'breakthrough' experiences.

When a psychedelic substance is ingested in a positive and supportive environment by a mentally sound individual (i.e., with positive set and setting), peak experiences can often be characterized by feelings of 'undifferentiated unity' or 'ego dissolution' (Lebedev et al., 2015; Pahnke & Richards, 1966). Ego-dissolution has been specifically defined as "a feeling that one's 'self,' 'ego,' or 'I' is disintegrating or that the border between one's self and the external world is dissolving" (Lebedev et al., 2015; p. 3137). Similar to mystical religious experiences (James, 1985), but in contrast to pathological experiences of depersonalization, these experiences are often perceived as being of a sacred, profound, and highly positive character (Griffiths et al., 2006; Pahnke & Richards, 1966), and have also been documented to result in lasting positive behavioral (Griffiths, Richards, Johnson, McCann, & Jesse, 2008) and personality (MacLean, Johnson, & Griffiths, 2011) changes (see below). The lasting value of peak experiences was also recognized in early psychedelic psychotherapy studies, which specifically sought to induce such mystical experiences through high-dose sessions in order to maximize long-term therapeutic value (Grof, 2008; Pahnke et al., 1970; Sherwood, Stolaroff, & Harman, 1962).

In a recent study, Lebedev and colleagues (2015) used fMRI to determine the neural activity that subserves the ego-dissolution state, as induced by psilocybin. Subjects were intravenously administered psilocybin prior to scanning, and following their emergence from the scanner they filled out a 24-item visual analogue scale (Lebedev et al., 2015). From these ratings, the researchers derived a principal component that specifically relates to ego-dissolution and used this in subsequent analyses.

The researchers found that ego-dissolution was associated with a decoupling of the medial temporal lobe from various brain networks (Lebedev et al., 2015). More specifically, ego-dissolution was correlated with reduced functional connectivity between the hippocampal formation and a sensorimotor network, frontoparietal control network, and salience network (Lebedev et al., 2015). Moreover, ego-dissolution was also highly correlated with a disruption of salience network integrity, and reduced interhemispheric communication (Lebedev et al., 2015). These results suggest that disruption of large-scale brain networks, and their connections in the psychedelic state, underlies the subjective experience of ego-dissolution.

A recent study of LSD also examined neural correlates of ego-dissolution (Carhart-Harris et al., 2016), with similar results. The authors first used seed-based analysis to examine resting-state functional connectivity between the parahippocampal cortex bilaterally and the rest of the brain. They found that decreased resting-state functional connectivity between the parahippocampal cortex and the retrosplenial cortex, two major hubs of the default mode network, predicted self-reported experiences of ego-dissolution (Carhart-Harris et al., 2016). They then examined global measures of network integration, finding that decreased default network integrity also significantly predicted self-reported ego-dissolution (Carhart-Harris et al., 2016). Collectively, these results suggest that compromised activity or connectivity within the default network, known to be critically involved in the sense of self (Northoff et al., 2006), correlates with the experience of the dissolution of self.

There are intriguing parallels here to dreaming and creative inspiration. As discussed above, historically speaking, true creativity and inspiration were often seen as a cooption and overshadowing of the individual self by divine or other 'higher' forces (McMahon, 2013). Although we know of no empirical research that has addressed this phenomenon in artists or other creative people, the intriguing combination of visionary experiences, ego-dissolution, and mystical union in artistic inspiration is reminiscent of peak psychedelic experiences and warrants further consideration and study.

Dreaming provides another interesting parallel. Aside from the parallel immersive visionary experience, dreaming is characterized by a particular disruption of the sense of self: access to autobiographical memory seems to be singularly abolished in most cases, leading to the loss of a coherent sense of who one is and a failure to recognize the incongruence and bizarreness of the dream experience. At the same time, however, a sense of self as an agent, moving through and acting in a world, remains intact. That is, even as the sense of self as an *agent* is maintained, the sense of an *autobiographical* self is lost. Whereas default network integrity is compromised during peak ego-dissolution experiences in psychedelic states (Carhart-Harris et al., 2016), and the default network appears to generally be deactivated during psychedelic experiences (Carhart-Harris, Erritzoe, et al., 2012; Deakin et al., 2008; Muthukumaraswamy et al., 2013), activation in numerous default network regions remains high during REM sleep and dreaming (Domhoff, 2011; Domhoff & Fox, 2015; Fox et al., 2013), potentially helping to explain the differential alterations in the sense of self. These contrasts across dream and psychedelic experience hint at the possibility that the 'self' might actually be decomposable into multiple components supported by different brain regions and networks (e.g. *agentive* vs. *autobiographical* senses of self) (Christoff, Cosmelli, Legrand, & Thompson, 2011).

*Resolution: Coming down*

As the acute psychedelic experience comes to an end, the user can feel generally sober and back in control, yet at the same time lingering effects can endure for prolonged periods (sometimes up to several hours or more). Cognition and perception may still be altered in subtle yet significant ways, and feelings of bliss, wellbeing, and gratitude or awe at the antecedent psychedelic experiences can be pervasive.

To our knowledge, no study has yet investigated the resolution stage of the psychedelic experience. Investigation of this stage faces some unique challenges. For instance, the uncertainty of the timing of its occurrence, as well as the difficulty in defining when the experience is really 'starting' to end, given that experiential intensity tends to rise and fall in waves (Fig. 1), are both problematic for empirical study. Nonetheless, the neural correlates of the resolution stage are of great interest because of the singular mix of lingering psychedelic effects combined with relative cognitive clarity and meta-awareness. We hope that future work can address this stage.

*Assimilation: Evaluating and integrating psychedelic experience*

As with the resolution stage, we know of no neuroimaging work that has directly addressed the stage we call 'assimilation,' i.e., evaluating and integrating the content of the preceding psychedelic experience. One study investigating ketamine found that acute decreases in default network connectivity persisted at a 24 hour follow-up fMRI scan, but no subjective reports were collected, precluding any speculation about the meaning of this effect for assimilation of the preceding ketamine experience (Scheidegger et al., 2012).

Some minimal questionnaire data has addressed subjective experiences 24 hrs after psychedelic substance administration and found that users reported a variety of lingering side effects, such as fatigue or a feeling of dreaminess, which were generally very mild (Studerus et al., 2011), but very little other data exists beyond anecdotal reports. One possibility is that evaluating the psychedelic experience may be akin to the subsequent evaluation of creatively-generated ideas (Ellamil et al., 2012), or interpreting and evaluating one's dreams the following morning – i.e., we expect executive resources would be recruited to evaluate the content and utility of experiences generated and perceived in altered states of consciousness. These possibilities remain to be studied in the future.

*Sequelae: Long-term effects on brain, behavior, and beliefs*

**Long-term effects on the brain.** Several long-term neurobiological effects have been reported following chronic exposure to psychedelic substances. Investigations in rats, for instance, have shown that chronic exposure to ketamine (Garcia et al., 2008) results in elevated levels of brain-derived neurotrophic factor (BDNF). These results are intriguing because BDNF plays a crucial role in facilitating neurogenesis, synaptogenesis, and neuron survival (Binder & Scharfman, 2004) – whereas low levels of BDNF are associated with numerous psychiatric and neurodegenerative disorders, such as depression (Brunoni, Lopes, & Fregni, 2008) and Alzheimer's disease (Phillips et al., 1991). Another study found no effect on BDNF levels, but did observe that chronic ketamine administration reversed the adrenal gland hypertrophy and elevated levels of stress hormones evoked in rats by a chronic mild stress paradigm (Garcia et al., 2009). These animal results at least present some intriguing possible neurophysiological correlates of the persisting psychological effects reported by human users (see below).

Most relevant to our discussion is a study that used morphometric neuroimaging to examine cortical thickness in chronic ayahuasca users (Bouso et al., 2015). The rationale of the study was that, because psychedelic substances appear to stimulate neurotrophic and transcription factors associated with synaptic plasticity (such as BDNF), large-scale changes might be visible in chronic users as investigated with morphometric neuroimaging. Compared to controls, Bouso and colleagues (2015) found that chronic ayahuasca users exhibited significant cortical thinning in the posterior cingulate cortex (among other regions) – a major hub of the default mode network strongly implicated in self-referential thinking (Andrews-Hanna et al., 2014; Buckner, Andrews-Hanna, & Schacter, 2008; Fox et al., 2015). Moreover, the extent of the thinning in posterior cingulate cortex was correlated with the duration and intensity of ayahuasca use, as well as self-reported feelings of self-transcendence (Bouso et al., 2015). Recall that, consistent with these results, deactivation and disintegration of the default network has repeatedly been found during psychedelic experience, especially in relation to depersonalization and ego-dissolution (Carhart-Harris, Erritzoe, et al., 2012; Deakin et al., 2008; Muthukumaraswamy et al., 2013).

An equivocal pattern of results was observed in executive brain areas, with significantly increased cortical thickness in the dorsal anterior cingulate cortex, but significant thinning in the dorsolateral prefrontal cortex (Bouso et al., 2015). Although this somewhat ambiguous pattern of results precludes any facile interpretation, it does provide preliminary evidence that morphometric neuroimaging might be useful in the study of long-term brain structure changes in chronic psychedelic users. In summary, although long-term neurobiological effects are supported by only very preliminary data, they nonetheless offer some tantalizing clues to the mechanisms that might account for the numerous long-term psychological effects observable after even a single high-dose session.

**Long-term psychological effects.** In contrast to the relative paucity of neurobiological data on the long-term sequelae of psychedelic experience, there are now several rigorous behavioral and questionnaire studies addressing its long-term psychological effects. The general conclusion of these investigations has been that even one or a few high-dose psychedelic experiences can have lasting and generally positive effects on attitudes, patterns of other drug use, and overall sense of wellbeing.

Early work had already suggested upon long-term (e.g., 6-month or 10-year) follow-up that LSD experiences had a number of lasting beneficial effects, such as decreased egocentricity; greater aesthetic appreciation of art, music, and nature; and more self-understanding (McGlothlin & Arnold, 1971; McGlothlin, Cohen, & McGlothlin, 1967). Recent work has corroborated and expanded upon these earlier studies.

For instance, a pioneering study by Griffiths and colleagues (2006) administered various doses of psilocybin to psychedelic-naïve participants and followed-up two months later with questionnaires about persisting behavior/attitude changes, as well as reports from community observers on changes in the users. Importantly, this study employed a double-blind design with an active control of methylphenidate, and control subjects also completed long-term follow-up questionnaires and were observed by community members. The authors found that at two-months' follow-up, as compared to methylphenidate, psilocybin was associated with significantly greater positive attitudes about life and/or oneself; positive mood changes; altruistic/positive social effects; positive behavior changes; and wellbeing and life satisfaction (Griffiths et al., 2006). Moreover, community observer ratings agreed with self-reports, in that significantly greater positive change was observed in the psilocybin vs. methylphenidate group (Griffiths et al., 2006). A subsequent follow-up at 14 months reported similarly positive results more than one year after the psychedelic experience (Griffiths et al., 2008). Recently, the same research group essentially replicated these findings with one- and 14-month follow-up in a separate sample (Griffiths et al., 2011).

Another recent study conducted long-term (8-16 months) follow-up with participants following 1-4 psilocybin sessions (Studerus et al., 2011). Whereas negative changes in beliefs and attitudes were hardly ever reported (between 1% and 7% of participants), many users (between 18% and 38%) reported positive changes in worldview, values, awareness of personal problems, relationship to one's body and other people, and relationship to the environment. Moreover, in a retrospective questionnaire, the original psychedelic experience was overwhelmingly described as 'enriching' and 'positive' (Studerus et al., 2011).

In summary, although more research clearly needs to be conducted with larger samples and with a wider variety of psychedelic substances, well-controlled and rigorous research currently suggests that psychedelic experience is associated with persistent and overwhelmingly positive changes in behaviors, attitudes, and sense of wellbeing. Conversely, there is negligible evidence that psychedelic substance use encourages increased consumption of other substances or leads to psychosis. A major task ahead will be developing an understanding of how these persistent psychological changes are mediated at the level of the brain.

**Conclusions and future directions**

The neural correlates of the psychedelic experience are increasingly being investigated with noninvasive functional neuroimaging methods. This research has now provided a preliminary understanding of the neural substrates of various stages and contents of psychedelic experience, elucidating how various stages differ from one another, and also relate to kindred 'altered' states of consciousness such as dreaming and creative thinking.

Several conclusions can be gleaned from the preceding review. First, psychedelic experiences involving strong visual hallucinatory components activate the same brain areas as 'natural' altered states involving high rates of visual imagery, most notably daydreaming and nighttime dreaming. Second, *peak* psychedelic experiences involving loss of the sense of the self or 'ego-dissolution' involve deactivation or disintegration of brain networks, most notably the DMN, that are widely thought to maintain and subserve an internal stream of thoughts and a coherent sense of self. Third, the effects of psychedelic experiences can be sustained long after the acute effects have worn off and substance concentration has reached zero. Potent psychedelic experiences appear to have long-term (potentially even lifelong) effects on brain, behavior, and beliefs. This drawn-out period of *assimilation* and *sequelae*, as we have termed them here, has parallels in dreaming and creative thinking. A compelling dream or a great creative discovery can be pondered and assimilated for months or years afterward, and can affect the life of the individual and others in profound ways.

These meager conclusions can represent only a first step on a long road leading toward understanding of the neural correlates of various stages of psychedelic experience and their relation to other altered states. Difficult but fascinating challenges lie ahead. From a theoretical point of view, a much more nuanced framework is required than the preliminary model proffered here – one that explicitly maps different stages of the psychedelic experience onto various stages of the creative process, as opposed to simply pointing toward intriguing similarities. From an empirical point of view, more work is needed to understand the neural correlates of the LSD experience, to follow up on a recent seminal study (Carhart-Harris et al., 2016). Similarly, neuroimaging of as-yet-uninvestigated psychedelic substances, such as mescaline and salvia divinorum (see Table 2), would be a welcome addition to the small but growing literature of the neural correlates of other major psychedelics. Other as-yet-unanswered empirical questions are broader: for instance, do psychedelics interact with and potentially facilitate the creative process through cognitive mechanisms (such as a broadening of attentional scope); via affective-motivational aspects, such as perseveration on a creative project; or some combination of these and other mechanisms? Finally, a major challenge for researchers will be to investigate these substances in a more natural fashion. Virtually every study to date has used intravenous injection of concentrated and pure forms of a given psychedelic substance (for a notable exception, however, see de Araujo and colleagues (2012), who used a natural ayahuasca brew). While this method provides obvious advantages to researchers and participants alike (e.g., allowing for precise dosage and rapid onset), intravenous injection is almost never the preferred route of administration for any psychedelic drug in either recreational or traditional settings. Although this method of delivery is of course a sensible alternative in these early days of renewed research, in order to increase ecological validity researchers will eventually need to tackle the challenges of investigating psychedelic experiences arising from orally-ingested, insufflated, or inhaled substances in their standard or natural (e.g., whole plant) forms.